# WHY LONG-TERM DEBT INSTRUMENTS CANNOT BE DEPOSIT SUBSTITUTES

*By Russell Stanley Q. Geronimo*[*]

## ABSTRACT

The definition of deposit substitutes in Philippine tax law fails to consider the maturity of a debt instrument. This makes it possible for long-term bonds to be considered as deposit substitutes if they meet the 20-lender rule, taxable at 20% final tax. However, long-term debt instruments cannot realistically function as deposit substitutes even if they fall in the hands of 20 or more lenders. *First*, long-term debt instruments cannot simultaneously replicate the twin features of capital preservation and liquidity, which are integral to the nature of a deposit substitute. *Second*, deposit substitutes are an integral part of the maturity transformation process (i.e. short-term borrowing for the purpose of long-term lending) in financial intermediaries, which means that they should have low borrowing cost, made possible only by having short-term maturity.

To prove these propositions, this paper situates the function of deposit substitutes within the context of shadow banking, where said instruments originated and are generally used. To show the incompatibility between a deposit substitute and a long-term debt instrument, the paper applies the fundamental theory of bond values to 10-year zero-coupon treasury notes called "PEACe Bonds" in *Banco De Oro, et al. vs. Republic* (2015 and 2016). The paper recommends that deposit substitutes should be limited to debt instruments with maturity of not more than 1 year.

## INTRODUCTION

The National Internal Revenue Code (NIRC) of 1997 provides the following definition for deposit substitutes:

> "[A]n alternative from of obtaining funds from the public (the term 'public' means borrowing from twenty [20] or more individual or corporate lenders at any one time) other than deposits, through the issuance, endorsement, or acceptance of debt instruments for the borrowers own account, for the purpose of relending or purchasing of receivables and other obligations, or financing their own needs or the needs of their agent or dealer."[1]

This definition fails to include one important element: the maturity of a debt instrument. This omission leads to absurd results, such as the possible characterization of a 10-year zero-coupon treasury note as a deposit substitute if it falls in the hands of 20 or more lenders. This is exemplified by the PEACe Bonds in *Banco De Oro, et al. vs. Republic* (G.R. No. 198756, January 13, 2015) and *Banco De Oro, et al. vs. Republic* (G.R. No. 198756, August 16, 2016). Under Section 22(Y) of the NIRC of 1997, it is a matter of indifference whether the principal amount is payable in 1 year, 5 years, or 30 years, as long as the bond or note meets the 20-lender rule.[2] However, there is simply no conceivable way that long-term debt instruments can function "like" a deposit. To call a 10-year debt instrument a "substitute" for deposit is a serious error in conceptualization, and revolts against basic principles of finance.[3]

---

[*] Juris Doctor, University of the Philippines – College of Law
[1] Sec. 22(Y), R.A. No. 8424

A deposit substitute is intended to be the economic equivalent of a bank deposit.[4] A debt instrument can only fulfill this role if it replicates or approximates a bank deposit's safety features.[5] These features are (i) capital preservation or safety of principal, and (ii) liquidity, if not withdrawability upon demand.[6] By nature, long-term debt instruments cannot do this, even if the issuer is the Republic of the Philippines or any sovereign entity.[7] A time to maturity of 10 years, coupled with lack of interest payments during the life of the instrument, exposes the bond to significant interest rate risk and price volatility.[8]

Bond prices fall if interest rates increase, and bond prices increase if interest rates decrease.[9] Hence, in an environment with rising interest rates, the value of a bond held by an investor may decrease below his initial investment.[10] The investor has two options: to sell the instrument prior to maturity at a loss, or to wait until maturity to redeem the instrument at par. If the investor chooses to sell at a loss, this negates the first safety feature of deposits: capital preservation. On the other hand, if he chooses to wait until maturity, this negates the second safety feature: liquidity. The fact that uncontrollable factors, like prevailing market interest rates, can adversely affect the performance of the investment makes a long-term debt instrument incompatible with the perceived "safety" of deposits.[11] Therefore, long-term debt instruments can never be deposit substitutes.

---

[2] G.R. No. 198756, January 13, 2015 ("Hence, when there are 20 or more lenders/investors in a transaction for a specific bond issue, the seller is required to withhold the 20% final income tax on the imputed interest income from the bonds.")

[3] Kores, L., *What is Shadow Banking?*, 50 FINANCE & DEVELOPMENT 2 (2013) ("Commercial banks engage in maturity transformation when they use deposits, which are normally short term, to fund loans that are longer term. Shadow banks do something similar. They raise [that is, mostly borrow] short-term funds in the money markets and use those funds to buy assets with longer-term maturities."); Borst, N., *Shadow Deposits as a Source of Financial Instability: Lessons from the American Experience for China*, 13 POLICY BRIEF PETERSON INSTITUTE FOR INTERNATIONAL ECONOMICS (2013) ("[S]hadow deposits are wealth management products, short-term financial products offered as an alternative to traditional savings deposits, most of which are principal unguaranteed and thus lack the implicit government support enjoyed by traditional deposits.")

[4] Whitehead, C.K., *Regulating for the Next Financial Crisis*, 24 PAC. MCGEORGE GLOBAL BUS. & DEV. LJ 3 (2011) ("[Money market mutual funds] and finance companies together provide the functional equivalent of deposit-taking and lending by banks.")

[5] Jackson, B.F., *Danger Lurking in the Shadows: Why Regulators Lack the Authority to Effectively Fight Contagion in the Shadow Banking System*, 127 HARVARD LAW REVIEW 2 (2013)

[6] *Id.*

[7] Litterman, R.B. and Iben, T., *Corporate Bond Valuation and the Term Structure of Credit Spreads*, 17 THE JOURNAL OF PORTFOLIO MANAGEMENT 3, 52-64 (1991)

[8] *Id.*

[9] Longstaff, F.A. and Schwartz, E.S., *Interest Rate Volatility and Bond Prices*, 49 FINANCIAL ANALYSTS JOURNAL 4, 70-74 (1993)

[10] *Id.*

[11] Malkiel, B.G., *Expectations, Bond Prices, and the Term Structure of Interest Rates*, 76 THE QUARTERLY JOURNAL OF ECONOMICS 2, 197-218 (1962)

Financial history supports this conclusion. Deposit substitutes are rooted in the shadow banking system.[12] "Shadow banks" are financial institutions that have bank-like functions but do not have the requisite license to operate as commercial banks, and are therefore not covered by bank regulation.[13] Like traditional banks, shadow banks engage in "maturity transformation", or the process of short-term borrowing for the purpose of long-term lending.[14]

Through maturity transformation, short-term liabilities in the form of "deposits" generate lesser interest expense, while long-term assets in the form of "loans" generate higher interest income, resulting in net interest income.[15] However, instead of borrowing from the public in the form of "deposits", shadow banks borrow in the form of "shadow deposits", which are also called "quasi-deposits", and which is just another term for "deposit substitutes".[16] And because shadow banks must keep the cost of borrowing low, it must necessarily resort to short-term borrowing.[17] This is why deposit substitutes necessarily have short-term maturity.[18]

The definition of deposit substitutes in the tax code ignores this context. It is therefore the task of this paper to illuminate this unexamined portion of the tax law, with a view toward a proper definition. The roadmap for discussion is as follows:

- **Part I** summarizes the tax implication of classifying a debt instrument as a deposit substitute under Section 22(Y) of the NIRC of 1997.

- **Part II** provides an overview of *Banco De Oro, et al. vs. Republic* (2015) and *Banco De Oro, et al. vs. Republic* (2016).

- **Part III** explains the underlying relationship between interest rates and maturity of debt instruments through the concept of a Yield Curve. This is necessary to understand why banks and financial institutions borrow in the short-term and lend in the long-term.

- **Part IV** explains the paradigmatic function of banking: short-term borrowing for the purpose of long-term lending, otherwise called as Maturity Transformation. This is important to understand the activity of shadow banks (which replicate the activity of banks) and the nature of deposit substitutes (which replicate the nature of deposits).

---

[12] Gorton, G. and Metrick, A., *Regulating the Shadow Banking System*, BROOKINGS PAPERS ON ECONOMIC ACTIVITY 2, 261-297 (2010)
[13] *Id.*
[14] Ricks, M., *Shadow Banking and Financial Regulation*, COLUMBIA LAW AND ECONOMICS WORKING PAPER NO. 370 (2014), available at: https://ssrn.com/abstract=1571290 (last accessed: 08 May 2017)
[15] Thakor, A.V., *Maturity Transformation*, THE NEW PALGRAVE DICTIONARY OF MONEY AND FINANCE 678-680 (1992)
[16] Lamberte, M., *Assessment of the Financial Market Reforms in the Philippines (1980-1992)*, 20 JOURNAL OF PHILIPPINE DEVELOPMENT 2, 231-59 (1993); Borst, N., *Shadow Deposits as a Source of Financial Instability: Lessons from the American Experience for China*, POLICY BRIEF PETERSON INSTITUTE FOR INTERNATIONAL ECONOMICS 13 (2013)
[17] *Supra* note 5.
[18] *Id.*

- **Part V** explains the features that make bank deposits relatively safe compared to other asset classes, and explains how banks provide safety to deposits.

- **Part VI** explains the shadow banking system, which is a set of practices that simulate traditional banking activity without bank regulation. This part also explains the origin of deposit substitutes, which are intended to simulate traditional bank deposits as a source of funding.

- **Part VII** explains how deposit substitutes replicate the safety of deposits without being governed by the prudential requirements imposed on banks.

- **Part VIII** explains why long-term debt instruments cannot function as substitutes for deposits through a discussion of the fundamental theory of bond values.

- **Part IX** applies the fundamental theory of bond values to the PEACe bonds, and demonstrate why 10-year zero-coupon treasury notes cannot function as deposit substitutes.

- **Part X** proposes a new definition for deposit substitutes by incorporating the concept of maturity.

## PART I: TAX IMPLICATION OF DEPOSIT SUBSTITUTES

The present definition of deposit substitutes in Section 22(Y) of the NIRC of 1997 is substantially the same as the definition in Section 95 of the New Central Bank Act. *First*, it is an alternative form of obtaining funds other than deposits. *Second*, the funds are obtained from the "public", which is defined as borrowing from 20 or more lenders at any one time. *Third*, the funds are obtained through the issuance, endorsement, or acceptance of debt instruments. *Fourth*, the use of funds is for the purpose of relending or purchasing of receivables and other obligations, or financing their own needs or the needs of an agent or dealer.[19]

The Manual of Regulation for Banks (MORB) recognizes the following examples of deposit substitutes: (i) repurchase agreements or "repos", (ii) promissory notes, (iii) certificates of participation with recourse, and (iv) certificates of assignment with recourse.[20]

The tax implication of classifying a debt instrument as a deposit substitute is illustrated as follows:

---

[19] Sec. 22(Y), R.A. No. 8424
[20] Bangko Sentral ng Pilipinas (BSP) Manual of Regulation for Banks (MORB) dated 31 October 2015

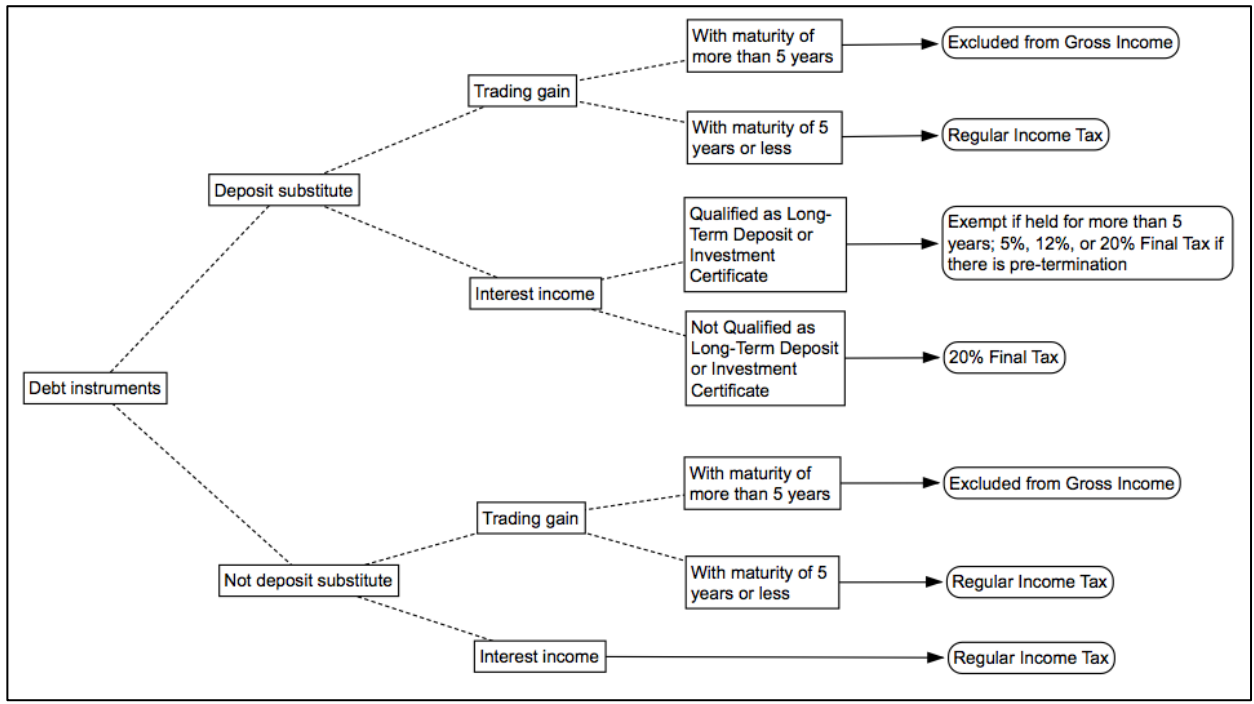

A debt instrument is an evidence of an obligation by a borrower to pay the lender the borrowed amount with interest at maturity. If it meets the elements under Section 22(Y), it is classified as a deposit substitute.

A debt instrument generates two types of gains: (i) trading gain and (ii) interest income. Trading gain pertains to the income by a holder of a debt instrument for selling the instrument prior to maturity, or for retiring the instrument at maturity. Interest income pertains to the gain for parting with the use of funds.

Regardless of whether a debt instrument is a deposit substitute or not, the treatment of trading gain is the same, pursuant to Section 32(B)(7)(g) of the NIRC, which provides:

> "Sec. 32. *Gross Income.* – […]
>
> (B) Exclusions from Gross Income. - The following items shall not be included in gross income and shall be exempt from taxation under this title: […]
>
> (7) Miscellaneous Items. – […]
>
> (g) Gains from the Sale of Bonds, Debentures or other Certificate of Indebtedness. - Gains realized from the sale or exchange or retirement of bonds, debentures or other certificate of indebtedness with a maturity of more than five (5) years."

The trading gain from debt instruments with maturity of more than 5 years is excluded from gross income. Otherwise, it is included in gross income and subject to the regular income tax rate.

Interest income on deposit substitutes has two possible tax treatments. It is generally subject to 20% final withholding tax, pursuant to Sections 24(B)(1), 25(A)(2), 27(D)(1), and 28(A)(7)(a). However, it is exempt from income taxation if the instrument is held for more than 5 years and classified as Long-Term Deposit or Investment Certificate under Section 22(FF) of the NIRC, which states:

> "(FF) The term *"long-term deposit or investment certificates"* shall refer to certificate of time deposit or investment in the form of savings, common or individual trust funds, deposit substitutes, investment management accounts and other investments with a maturity period of not less than five (5) years, the form of which shall be prescribed by the Bangko Sentral ng Pilipinas (BSP) and issued by banks only (not by nonbank financial intermediaries and finance companies) to individuals in denominations of Ten thousand pesos (P10,000) and other denominations as may be prescribed by the BS."

If the taxpayer pre-terminates the instrument, the interest income is subject to 5%, 12%, or 20% depending on the length of time when the instrument was held.

Finally, interest income on debt instruments that are not deposit substitutes are not subject to 20% final withholding tax, but to the regular income tax rate.

The tax implication of classifying a debt instrument as deposit substitute has drastic economic consequences. It can mean billions of potential tax revenues, or potential tax liability. This is exemplified by the landmark decisions of *Banco De Oro, et al. vs. Republic* (G.R. No. 198756, January 13, 2015) and *Banco De Oro, et al. vs. Republic* (G.R. No. 198756, August 16, 2016) on PEACe Bonds, discussed in the next section.

## PART II: OVERVIEW OF *BANCO DE ORO, ET AL. VS. REPUBLIC*

On 23 March 2001, the Caucus of Development NGO Networks (CODE-NGO) requested the Department of Finance to issue 10-year zero-coupon Treasury Certificates through the Bureau of Treasury. The CODE-NGO proposed to purchase the Certificates through a special purpose vehicle, which in turn will re-package the Certificates as PEACe Bonds to be sold at a premium to investors. The proceeds of the bond issuance will be placed in the Hanapbuhay Fund, which is a permanent fund to finance the projects of accredited Non-Government Organizations (NGOs).

On 31 May 2001, the BIR issued Ruling No. 020-2001, which states that the 10-year zero coupon bonds are not deposit substitutes, and therefore not subject to 20% final withholding tax, because they will be originally issued to only one entity: CODE-NGO. To be a deposit substitute, the funds must be obtained from twenty (20) or more borrowers.

Subsequently, the BIR issued Ruling No. 035-2001 dated 16 August 2011 and Ruling No. DA-175-01 dated 29 September 2001, both reiterating that the 10-year zero-coupon bonds are not deposit substitutes for failing to meet the 20-lender rule. Specifically, the phrase "at any one time" is to be determined at the time of the original issuance of the bonds.

The Bureau of Treasury decided not to issue the bonds to a special purpose vehicle of CODE-NGO because it is not a Government Securities Eligible Dealer (GSED). Hence, on 09 October 2001, it issued a Notice of Public Offering of Treasury Bonds to all GSEDs, stating that P30 billion worth of 10-year zero-coupon bonds will be auctioned on 16 October 2001. The Notice specifically provides that the bonds will not be issued to more than 19 buyers or lenders.

On 16 October 2001, the Bureau of Treasury held the auction for the 10-year zero-coupon bonds. RCBC Capital, which participated on behalf of CODE-NGO, was declared as the winning bidder. On the same date, RCBC Capital and CODE-NGO executed an Underwriting Agreement, with RCBC Capital as the Issue Manager and Lead Underwriter for the offering of the PEACe Bonds.

On 18 October 2001, the Bureau of Treasury issued P35 billion worth of zero-coupon bonds to RCBC, with yield-to-maturity of 12.75%, for approximately P10.17 billion, resulting in a discount of approximately P24.83 billion.

Subsequently, RCBC Capital sold the bonds in the secondary market for an issue price of approximately P11.99 billion. The purchasers include Banco de Oro, Bank of Commerce, China Banking Corporation, Metropolitan Bank & Trust Company, Philippine Bank of Communications, Philippine National Bank, Philippine Veterans Bank, and Planters Development Bank.

On 07 October 2011, the BIR issued Ruling No. 370-2011, which states that PEACe Bonds are deposit substitutes, and therefore the discount of P24.83 billion is treated as interest income subject to 20% final withholding tax, pursuant to Section 27(D)(1) of the NIRC of 1997. Pursuant to the ruling, the Secretary of Finance directed the Bureau of Treasury to withhold the tax from the face value of the PEACe Bonds upon payment at maturity. As a consequence, the holders of the PEACe Bonds filed a petition for certiorari, prohibition and/or mandamus before the Supreme Court.

The issue before the Court was whether the PEACe Bonds are deposit substitutes. In the Decision dated January 13, 2015, the Court ruled as follows:

> "Applying Section 22(Y) of the National Internal Revenue Code, we held that the number of lenders/investors at every transaction is determinative of whether a debt instrument is a deposit substitute subject to 20% final withholding tax. When at any transaction, funds are simultaneously obtained from 20 or more lenders/investors, there is deemed to be a public borrowing and the bonds at that point in time are deemed deposit substitutes. Consequently, the seller is required to withhold the 20% final withholding tax on the imputed interest income from the bonds. We further declared void BIR Rulings Nos. 370-2011 and DA 378-2011 for having disregarded the 20-lender rule provided in Section 22(Y)."

The Court issued a Resolution in August 16, 2016, affirming the 2015 Decision on the nature of deposit substitutes:

> "[…] in light of Section 22(Y), the reckoning of whether there are 20 or more individuals or corporate lenders is crucial in determining the tax treatment of the yield from the debt instrument. In other words, if there are 20 or more lenders, the debt instrument is considered a *deposit substitute* and subject to 20% final withholding tax."

As will be discussed, however, the 20-lender rule should not be the only determining factor for classifying a debt instrument as a deposit substitute. The maturity of the instrument is also essential.

**PART III: THE YIELD CURVE**

The elements of a debt instrument are: (i) principal amount, (ii) coupon, (iii) maturity, (iv) yield, (v) time to maturity, and (vi) coupon structure. The principal amount is the amount borrowed by the issuer.[21] The coupon is the consideration paid to the lender for lending the principal amount.[22] The maturity is the date when the lender is obliged to repay the principal amount.[23] The yield is the rate of return on the investment.[24] The time to maturity is the remaining time before the principal amount is repaid.[25] The coupon structure is the timing and frequency of coupon payments.[26]

The yield curve is a graph that describes the relationship between the yield and the time to maturity.[27] It is represented by an upward-sloping curve, which indicates that the longer is the time to maturity, the higher is the yield.[28] The following is an example of a yield curve for government bonds in the Philippines[29]:

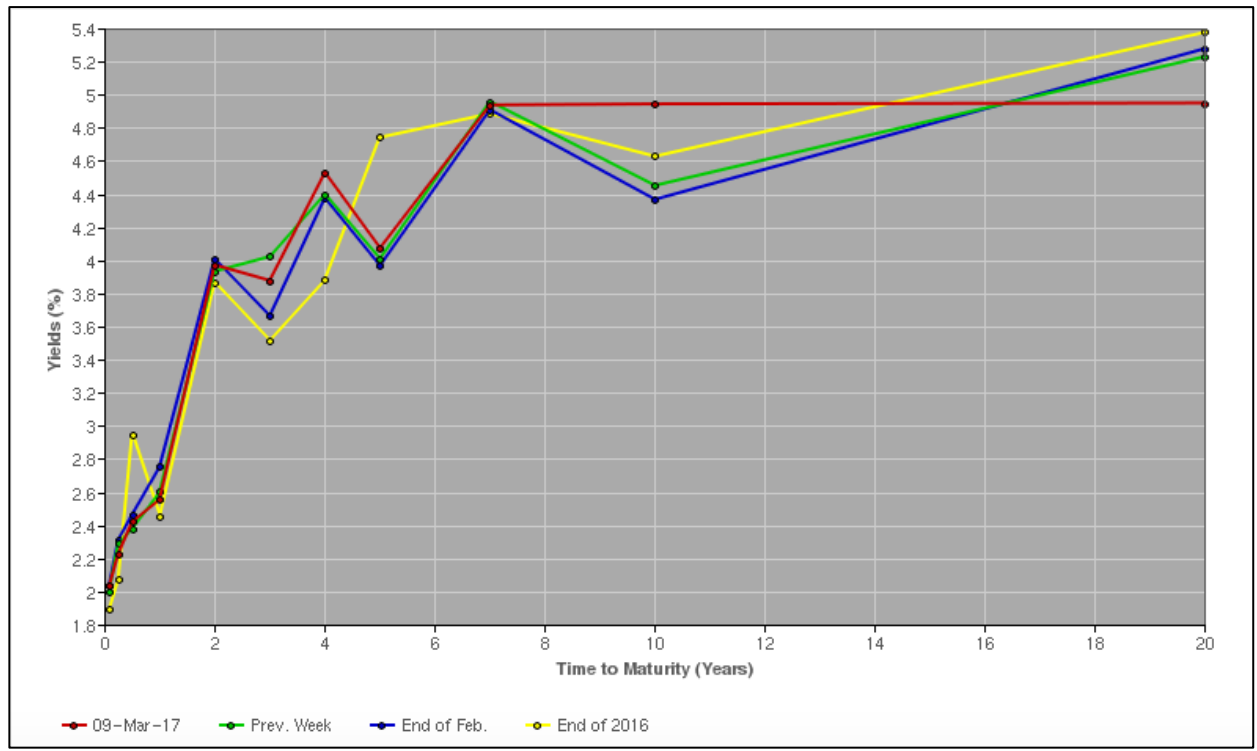

The shape of the yield curve is generally upward-sloping.[30] One explanation for this is the liquidity premium theory, which states that investors prefer highly liquid and short-term assets, and therefore investors demand a higher return on investment from a long-term debt instrument to compensate them for the risk of giving up liquidity for a longer period.[31] A long time to maturity exposes the investment to several risks, such as the possibility of default, changes in interest rates, and changes in the price of the instrument.[32] This theory merely restates one of the central principles of investment: that the greater is the risk assumed by the investor, the greater is the expected return.[33] This is important to understand the process of maturity transformation in banks and non-bank financial intermediaries.

---

[21] *Supra* note 7.
[22] *Id.*
[23] *Id.*
[24] *Id.*
[25] *Id.*
[26] *Id.*
[27] Campbell, J.Y., *Some Lessons from the Yield Curve*, 9 THE JOURNAL OF ECONOMIC PERSPECTIVES 3, 129-152 (1995)
[28] Barnes, T., and Burnie, D.A., *The Estimation of Corporate Bond Yield Curves as a Function of Term to Maturity and Coupon*, 26 QUARTERLY JOURNAL OF BUSINESS AND ECONOMICS 4, 50-64 (1987)
[29] Government Bond Yield Curves for Selected Periods, ASIANBONDSONLINE, available at:
 https://asianbondsonline.adb.org/philippines/data/marketwatch.php?code=government_bond_yields  (last retrieved on 05 April 2017)
[30] Worley, R.B., and Diller, S., *Interpreting the Yield Curve*, 32 FINANCIAL ANALYSTS JOURNAL 6, 37-45 (1976)

**PART IV: MATURITY TRANSFORMATION**

Maturity transformation is the process of short-term borrowing for the purpose of long-term lending.[34] As shown in Part III, short-term debt instruments generally have lower yields compared to long-term debt instruments.[35] To make a profit, a bank must issue short-term debt instruments and invest in long-term assets.[36] Issuing short-term debt instruments entail low borrowing cost, while investing in long-term debt instruments entail higher yields.[37]

A bank deposit is technically a short-term debt instrument.[38] It is short-term because a deposit is withdrawable by the depositor upon demand.[39] Of course, the depositor may choose not to withdraw the deposit for a long period of time, but this does not affect the short-term character of the instrument.[40]

In the Philippines, deposit interest rates have an average of 9.13% from CY 1980 to CY 2015.[41] The all-time high rate is 21.17%, which happened in CY 1984.[42] The all-time low is 1.23% in CY 2014.[43] It hit 1.6% in CY 2015. The following shows the trend of deposit interest rates:[44]

---

[31] Van Horne, J., *Liquidity Premiums and the Government Bond Market*, 20 FINANCIAL ANALYSTS JOURNAL 5, 127-129 (1964)
[32] Kiely, J.K., Kolari, J.W., and Rose, P.S., *A Reexaminaron of the Relationship between Liquidity Premiums and the Level of Interest Rates*, 34 QUARTERLY JOURNAL OF BUSINESS AND ECONOMICS 3, 60-70 (1995)
[33] Campbell, J.Y., *Understanding Risk and Return*, 104 JOURNAL OF POLITICAL ECONOMY 2, 298-345 (1996)
[34] Niinimaki, J., *Maturity Transformation without Maturity Mismatch and Bank Panics*, 159 JOURNAL OF INSTITUTIONAL AND THEORETICAL ECONOMICS 3, 511-522 (2003)
[35] Chun, A.L., *Expectations, Bond Yields, and Monetary Policy*, 24 THE REVIEW OF FINANCIAL STUDIES 1, 208-247 (2011)
[36] Kashyap A.K., Rajan, R., and Stein, J.C., *Banks as Liquidity Providers: An Explanation for the Coexistence of Lending and Deposit-Taking*, 57 THE JOURNAL OF FINANCE 1, 33-73 (2002)
[37] Bond, P., *Bank and Nonbank Financial Intermediation*, 59 THE JOURNAL OF FINANCE 6, 2489-2529 (2004)
[38] Cannan, E., *The Meaning of Bank Deposits*, ECONOMICA 1, 28-36 (1921)
[39] Cramp, A.B., *Financial Theory and Control of Bank Deposits*, 20 OXFORD ECONOMIC PAPERS 1, 98-108 (1968)
[40] Gray, H.P., *Higher Interest Rates on Time Deposits: Reply*, 20 THE JOURNAL OF FINANCE 1, 86-88 (1965)
[41] Deposit Interest Rate in the Philippines, TRADING ECONOMICS, available at:
http://www.tradingeconomics.com/philippines/deposit-interest-rate (last retrieved on 05 March 2017)
[42] *Id.*
[43] *Id.*
[44] *Id.*

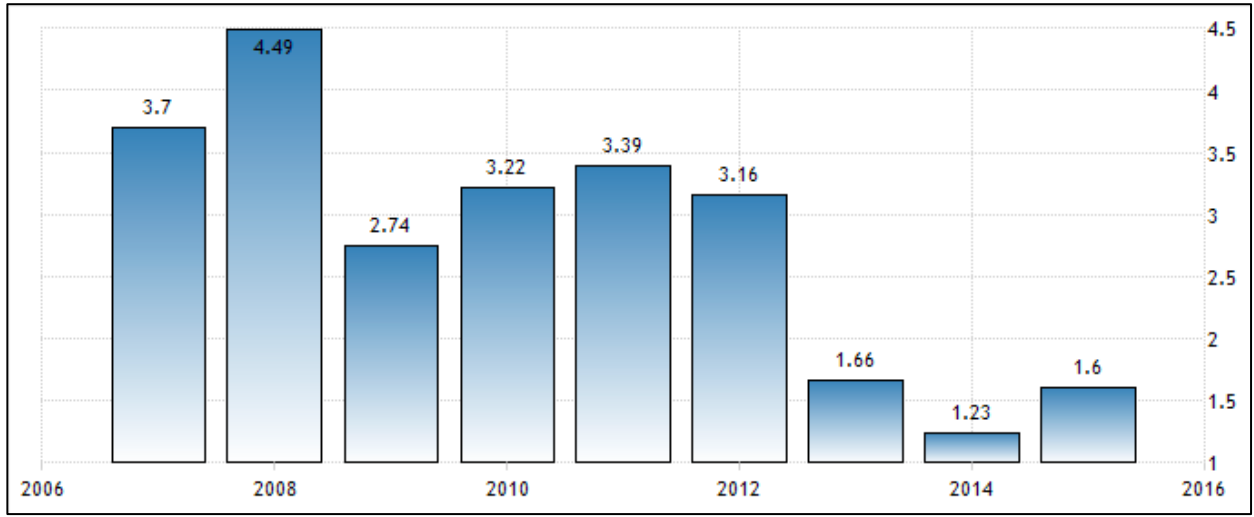

Loans, which are funded by bank deposits, are typically long-term in character, and therefore entail higher returns.[45] Bank lending rates in the Philippines have an average of 13.45% from CY 1976 to CY 2016, with an all-time high of 39.73% in CY 1984 and an all-time low of 5.09% in CY 2015.[46] The following is the trend of bank lending rates:[47]

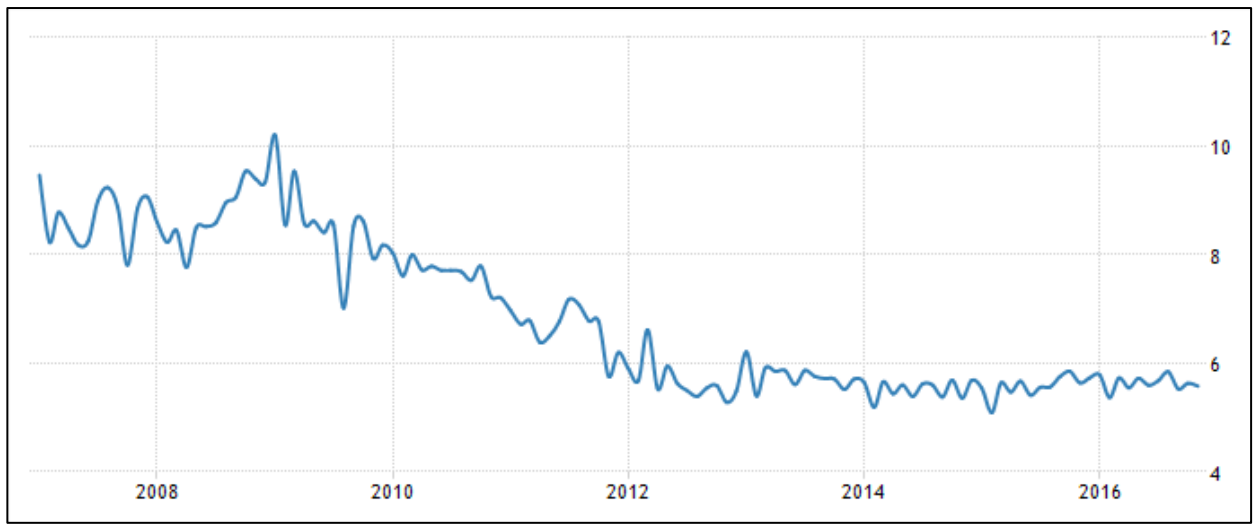

Deposits are liabilities of the bank, while loans are assets.[48] Interest expense on deposits represents the cost of borrowing, while interest income on the loans represents return on investment.[49] The difference between the cost of borrowing and the return on

---

[45] Stone, B.K., *The Cost of Bank Loans*, 7 THE JOURNAL OF FINANCIAL AND QUANTITATIVE ANALYSIS 5, 2077-2086 (1972)
[46] Philippines Bank Lending Rate, TRADING ECONOMICS, available at: http://www.tradingeconomics.com/philippines/bank-lending-rate (last retrieved on 05 March 2017)
[47] *Id.*
[48] Stein, J.C., *Securitization, Shadow Banking & Financial Fragility*, 139 DAEDALUS 4, 41-51 (2010)
[49] *Id.*

investment is called the spread.[50] The wider is the spread, the higher is the profit of the bank.[51]

**PART V: SAFETY FEATURES OF TRADITIONAL BANK DEPOSITS**

A bank deposit is a special form of debt instrument, described as follows:

> "First, they are demand-debt instruments, which means they give depositors the legal right to withdraw their money at will. Second, they give depositors the ability to withdraw their money at virtually no cost. And third, they are subject to only negligible interest rate risk."[52]

In short, the first safety feature is withdrawability upon demand or liquidity, and the second is safety of principal. The third feature is only a restatement of the second—i.e. interest rate risk only affects the price of the instrument, which in turn affects the safety of principal.[53] As will be explained in Part VIII, the relationship between the price of a debt instrument and prevailing market interest rates is inverse.[54] Hence, a negligible interest rate risk only means that there is very low probability of devaluation in the value of a deposit.[55]

The perceived safety of bank deposits is due to bank regulation, which imposes the following prudential requirements or limitations: (i) capital requirements, (ii) deposit insurance, (iii) reserve requirements, (iv) government liquidity facilities, (v) periodic supervisory examinations, (vi) Single Borrower's Limit, (vii) Directors, Officers, Stockholders and their Related Interests (DOSRI) Limits, (viii) investment limitations, and (ix) other prudential regulatory requirements.[56]

Implementing these strict regulatory measures to protect bank deposits is an additional economic burden on banks.[57] These measures constitute huge regulatory costs in doing business.[58] Meanwhile, no similar measures were initially imposed on money market mutual funds and other non-bank financial institutions.[59] This explains the eventual rise of deposit substitutes.[60]

There are three major protections that deposit substitutes lack: (i) mandatory deposit insurance coverage, (ii) reserve requirements, and (iii) stringent capital requirements.[61]

---

[50] Krishna, C.N.V., Ritchken, P.H., and Thomson, J.B., *On Credit-Spread Slopes and Predicting Bank Risk*, 38 JOURNAL OF MONEY, CREDIT AND BANKING 6, 1545-1574 (2006)
[51] *Id.*
[52] *Supra* note 5.
[53] Craine, R.N., and Pierce, J.L., *Interest Rate Risk*, 13 THE JOURNAL OF FINANCIAL AND QUANTITATIVE ANALYSIS 4, 719-732 (1978)
[54] Brennan, M.J., and Schwartz, E.S., *Bond Pricing and Market Efficiency*, 38 FINANCIAL ANALYSTS JOURNAL 5, 49-56 (1982)
[55] *Id.*
[56] R.A. No. 8791 (The General Banking Law of 2000)
[57] *Supra* note 5.
[58] *Id.*
[59] *Supra* note 12.
[60] *Id.*
[61] Levitin, A.J., *Safe Banking: Finance and Democracy*, 83 THE UNIVERSITY OF CHICAGO LAW REVIEW 1, 357-455 (2016)

## A. Mandatory Deposit Insurance Coverage

Bank deposits are required to be covered by a state deposit insurance system.[62] This serves to protect depositors from the risk of non-payment of deposits in case the bank becomes insolvent.[63] The impetus for this coverage is again the Great Depression, when there were several bank runs and depositors not getting their deposits back.[64] The deposit insurance system obligates banks to pay a certain percentage of their deposits, in the form of premiums, to a deposit insurer.[65] When the bank is closed down by the monetary board, the deposit insurer indemnifies the depositors to the extent that they are unable to immediately recover from the bank.[66] The indemnification, however, is only up to a certain limit. The portion of the deposit not covered by the insurance may still be recovered from the assets of the bank upon dissolution and liquidation.[67]

## B. Reserve Requirements

Reserve requirements or cash reserve ratio obligates banks to set aside a fraction of deposits which they cannot lend out to the borrowing public.[68] The bank either retains the bank deposits in the vault or further deposits them with the central bank.[69] The reserve requirement is one of the monetary tools of the central bank in influencing the level of interest rates.[70]

## C. Stringent Capital Requirements

Banking institutions are required to have capital adequacy through the imposition and observance of minimum capital requirements.[71] Today, the global standard regulatory framework for capital adequacy is the Third Basel Accord or Basel III, which provides that banks must have common equity of 4.5% of risk-weighted assets.[72]

---

[62] Leaven, L., *Bank Risk and Deposit Insurance*, 16 THE WORLD BANK ECONOMIC REVIEW 1, 109-137 (2002)
[63] Wheelock, D.C., and Wilson, P.W., *Explaining Bank Failures: Deposit Insurance, Regulation, and Efficiency*, 77 THE REVIEW OF ECONOMICS AND STATISTICS 4, 689-700
[64] Wagster, J.D., *Wealth and Risk Effects of Adopting Deposit Insurance in Canada: Evidence of Risk Shifting by Banks and Trust Companies*, 39 JOURNAL OF MONEY, CREDIT AND BANKING 7, 1651-1681 (2007)
[65] Cooper, R., and Ross, T.W., *Bank Runs: Deposit Insurance and Capital Requirements*, 43 INTERNATIONAL ECONOMIC REVIEW 1, 55-72 (2002)
[66] Cull, R., Senbet, L.W., and Sorge, M., *Deposit Insurance and Financial Development*, 37 JOURNAL OF MONEY, CREDIT AND BANKING 1, 43-82 (2005)
[67] *See*, e.g., R.A. No. 3591 as amended
[68] Section 94, R.A. No. 7653
[69] Gambs, C.M., *State Reserve Requirements and Bank Cash Assets*, 12 JOURNAL OF MONEY, CREDIT AND BANKING 3 (1980)
[70] Baltensperger, E., *Reserve Requirements and Economic Stability*, 14 JOURNAL OF MONEY, CREDIT AND BANKING 2 (1982)
[71] Diamond, D.W., and Rajan, R.G., *A Theory of Bank Capital*, 55 THE JOURNAL OF FINANCE 6 (2000)
[72] Hartlage, A.W., *The Basel III Liquidity Coverage Ratio and Financial Stability*, 111 MICHIGAN LAW REVIEW 3 (2012)

# PART VI: THE RISE OF DEPOSIT SUBSTITUTES THROUGH THE SHADOW BANKING SYSTEM

Shadow banks are non-bank financial institutions which mirror the maturity transformation process in banks.[73] Both traditional banks and shadow banks issue short-term debt instruments as a source of funding—i.e. deposits in the case of banks and deposit substitutes in the case of shadow banks.[74] They both invest in long-term debt instruments—i.e. loans in the case of banks, and receivables, bonds, and notes in the case of shadow banks.[75] In short, shadow banks are direct competitors of banks, and shadow banks have a competitive advantage because they are not covered by bank regulation.[76] Historically, a lax regulatory environment allowed shadow banks to expand operations faster than licensed banks.[77]

Over time, however, laws were introduced allowing traditional banks to do whatever shadow banks were doing.[78] This is the reason why universal and commercial banks are now allowed to have "quasi-banking" functions.[79] This is the reason why banks do not merely issue deposits, but also "quasi-deposits".[80] Quasi-banking is just another term for shadow banking, and quasi-deposit is just another term for deposit substitute.[81]

## A. Money Market Funds as the First Shadow Banks

The issuance of quasi-deposits or deposit substitutes was initially done by money market funds.[82] Through a money market fund, a non-bank financial institution accepts cash from the public, which is used by the institution to purchase short-term debt securities, such as treasury bills and commercial papers.[83] The clients are technically "lenders" and the non-bank financial institutions are "borrowers".[84] The borrowers oblige themselves to pay a fixed rate of interest to the lenders.[85] They profit from the difference between the yields on the investment instruments and the interest rate obligation.[86] The first money market fund in the U.S. is The Reserve Fund, which was established in 1971 by Bruce R. Bent and Henry B.R. Brown.[87]

---

[73] Gennaioli, N., Shleifer, A., and Vishny, R.W., *A Model of Shadow Banking*, 68 THE JOURNAL OF FINANCE 4 (2013)
[74] *Supra* note 61.
[75] *Id.*
[76] *Supra* note 5.
[77] *Id.*
[78] *See*, e.g., Depository Institutions Deregulation and Monetary Control Act of 1980 (H.R. 4986, Pub.L. 96–221)
[79] Section 5, R.A. No. 8791
[80] Section 5, R.A. No. 8791
[81] Section 249, P.D. No. 69
[82] Pozsar, Z., Adrian, T., Ashcraft, A.B., and Boesky, H., *Shadow Banking* (2010), available at SSRN: https://ssrn.com/abstract=1640545 or http://dx.doi.org/10.2139/ssrn.1640545 (last retrieved on 05 March 2017)
[83] *Couldock Bohan v. Societe Generale Sec. Corp.*, 93 F. Supp.2d 220 (D. Conn. 2000)
[84] *In re Cormack*, 124 B.R. 806 (Bankr. C.D. Cal. 1991)
[85] *Tibble V. Edison International*, CV 07-5359 SVW (AGRx) (C.D. Cal. Jul. 8, 2010)
[86] *Paine, Webber V. Merrill Lynch, Pierce*, 564 F. Supp. 1358 (D. Del. 1983)
[87] Finkle, V., *Shadow Banking* (2017), available at: http://businessresearcher.sagepub.com/sbr-1863-101611-2765611/20170102/shadow-banking (last retrieved on 05 March 2017)

## B. Expansion of the Shadow Banking Market

Four legislative acts in the U.S. enabled the expansion of the shadow banking market: (i) the McFadden Act of 1927, (ii) the Banking Act of 1933 (which spawned an important regulatory rule called "Regulation Q"), (iii) the Investment Company Act of 1940, and (iv) the Bank Holding Company Act of 1956.

### 1. McFadden Act of 1927

The McFadden Act of 1927 placed a restriction on interstate branching among banks in the U.S. Each national bank could only put up branches within the state where it is located.[88] This limitation on branch banking gave a competitive advantage to money market funds, which were not governed by a similar prohibition.[89] Compared to federal commercial banks, the money market funds could freely expand branch operations across different states.[90]

### 2. Banking Act of 1933 and Regulation Q

The Banking Act of 1933, which is otherwise known as the Glass-Steagall Act, introduced a comprehensive and nationwide banking reform to ensure the safety of bank deposits.[91] Section 11 of the Act and Title 12, part 217 of the United States Code of Federal Regulations introduced Regulation Q, which prohibited commercial banks from offering interest on checking accounts and regulated the level of interest rates on other deposit accounts, such as savings accounts, time deposits, and eventually Negotiable Order of Withdrawal accounts.[92] It placed a ceiling on the allowable rates, which can be adjusted by the central bank from time to time.[93]

The concept of placing an interest rate ceiling is based on the experience during the Great Depression in the 1930s, when banks competed freely in obtaining deposits from the public.[94] This was done by offering higher interest rates on deposit.[95] Since increasing the bank's interest obligation diminishes the margin between borrowing rates and lending rates, the bank engaged in speculative investments in order to maintain or increase their profitability.[96] Since higher interest rates on deposits meant higher interest expenditures, banks must match them with assets having higher yields, which meant higher risks.[97] This created perverse incentives to enter into risky and speculative loans.[98] Since the borrowing clients of banks have higher probability of default, the bank is exposed to higher risk of bankruptcy.[99]

---

[88] Luttrell, D., Rosenblum, H., and Thies, J., *Understanding the Risks Inherent in Shadow Banking: A Primer and Practical Lessons Learned*, STAFF PAPERS 18, Federal Reserve Bank of Dallas (2012)
[89] *Id.*
[90] *Id.*
[91] Preston, H.H., *The Banking Act of 1933*, 23 THE AMERICAN ECONOMIC REVIEW 4 (1933)
[92] Freidman, B.M., *Regulation Q and the Commercial Loan Market in the 1960s*, 7 JOURNAL OF MONEY, CREDIT AND BANKING 3 (1975)
[93] *Id.*
[94] Calomiris, C.W., *The Political Lessons of Depression-Era Banking Reform*, 26 OXFORD REVIEW OF ECONOMIC POLICY 3 (2010)
[95] Mayer, T., and Nathan, H., *Mortgage Rates and Regulation Q: Note*, 15 JOURNAL OF MONEY, CREDIT AND BANKING 1 (1983)
[96] *Id.*

In this environment, banks were vulnerable from becoming illiquid and insolvent, which prejudiced the depositors whose only security is limited to the assets of the bank upon dissolution.[100] Regulation Q, in addition to the introduction of mandatory deposit insurance under the Federal Deposit Insurance Corporation, was meant to arrest the free competition on bank deposits in order to eliminate these perverse incentives.[101]

Regulation Q played an important role in the growth of shadow deposits.[102] During the 1970s, there was a rapid increase in the general price of goods and services in the U.S. To arrest this inflation spike, the government raised interest rates to restrict credit and to slow down the economy. This was done through the Federal Reserve Bank, which raised higher interest rates on treasury bills and other short-term government debt securities. Higher interest rates led to a higher savings rate, which decreased consumer expenditures, which in turn limited the growth of money supply, and which would finally curb inflation.[103]

Meanwhile, the interest on bank deposit accounts was capped by Regulation Q.[104] Eventually, yields on short-term government debt securities exceeded the interest rate ceiling on bank deposits.[105] Since the former are safer investments than the latter, and yielded higher interest, the lending public shifted their savings from banks to money market funds, which institutionalized the purchase of government securities.[106]

### 3. Investment Company Act of 1940

The Investment Company Act of 1940 provided the regulatory framework for money market funds, which were incorporated under the Act as investment companies.[107] The Act classified these companies into three: face-amount certificate companies, unit investment trusts, and management companies. Face-amount certificate companies issue debt instruments that are repayable at par. Unit investment trusts issue redeemable securities representing participation rights in a basket of financial assets. They are established through agency, custodianship or trust indenture. Management companies are those that are neither face-amount certificate companies nor unit investment trusts.

---

[97] Lam, C.H., and Chen, A.H., *Joint Effects of Interest Rate Deregulation and Capital Requirements on Optimal Bank Portfolio Adjustments*, 40 THE JOURNAL OF FINANCE 2 (1985)
[98] *Id.*
[99] *Id.*
[100] Kane, E.J., *Getting Along Without Regulation Q: Testing the Standard View of Deposit-Rate Competition During the "Wild-Card Experience"*, 33 THE JOURNAL OF FINANCE 3 (1978)
[101] *Supra* note 92.
[102] Duca, J.V., *How Capital Regulation and Other Factors Drive Shadow Banking in the Short and Long Run*, available at: http://dev-som.yale.edu/sites/default/files/files/ShadowBanking_April_Yale_2014-1398177745_9471(1).pdf (last retrieved on 05 March 2017)
[103] *Id.*
[104] *Supra* note 97.
[105] *Supra* note 95.
[106] *Supra* note 92.
[107] Bosland, C.C., *The Investment Company Act of 1940 and Its Background*, 49 JOURNAL OF POLITICAL ECONOMY 4 (1941)

The main thrust of the Act is to build public confidence in holding securities.[108] These introduced new investment vehicles other than deposits in commercial banks.

### 4. Bank Holding Company Act of 1956

Lastly, the Bank Holding Company Act of 1956 prohibited a bank holding company in one state to acquire ownership interest in another bank located in another state, including interstate mergers.[109] No similar special prohibition governs money market funds.

### C. How Traditional Banks Absorbed Shadow Banking Functions

There were three legislative acts in the U.S. that tried to eliminate the competitive advantage of money market funds over commercial banks: (i) the Depository Institutions Deregulation and Monetary Control Act of 1980, (ii) the Garn-St. Germain Depository Institutions Act of 1982, and (iii) the Riegle-Neal Interstate Banking and Branching Efficiency Act of 1994.

These Acts further contributed to the growth of shadow banking by allowing commercial banks to compete with money market funds by issuing shadow deposits themselves. There were therefore two pathways for the issuance of shadow deposits: commercial banks and non-bank financial institutions. In the Philippines, this is equivalent to the co-existence of banks and non-banking financial institutions both having quasi-banking functions.

### 1. Depository Institutions Deregulation and Monetary Control Act of 1980

The Depository Institutions Deregulation and Monetary Control Act of 1980 expanded the control of the Federal Reserve to cover more types of banks.[110] Credit unions and savings and loan associations were authorized to issue demand deposits.[111] The Act also lifted the prohibition on bank mergers.[112] More importantly, it eliminated the authority of the Federal Reserve Board to prescribe interest rate ceilings on bank deposits, effectively repealing Regulation Q under the Glass-Steagall Act, with a staggered implementing period of 6 years.[113] The Act also legitimized the offering of Negotiable Order of Withdrawal accounts and increased the coverage of mandatory deposit insurance.[114] All these factors had the effect of lifting the disadvantages of banks compared to money market funds.[115]

---

[108] *Id.*
[109] Klebaner, B.J., *The Bank Holding Company Act of 1956*, 24 SOUTHERN ECONOMIC JOURNAL 3 (1958)
[110] Gilbert, R.A., *Requiem for Regulation Q: What It Did and Why It Passed Away, Federal Reserve Bank of St. Louis* (1986), available at:
 https://files.stlouisfed.org/files/htdocs/publications/review/86/02/Requiem_Feb1986.pdf (last retrieved on 05 March 2017)
[111] *Id.*
[112] *Id.*
[113] *Id.*
[114] *Id.*
[115] *Id.*

## 2. Garn-St. Germain Depository Institutions Act of 1982

The Garn-St. Germain Depository Institutions Act of 1982 allowed commercial banks to offer money market deposit accounts and Super Negotiable Orders of Withdrawal.[116] These products directly competed with the money market accounts offered by money market mutual funds.

## 3. Riegle-Neal Interstate Banking and Branching Efficiency Act of 1994

Riegel-Neal Interstate Banking and Branching Efficiency Act of 1994 allowed federally chartered banks to compete with state-chartered banks, and removed prohibitions on interstate branch banking.

## C. Shadow Banking in the Philippines

Shadow banking is legitimized in the Philippine financial system through the issuance of quasi-banking licenses by the Bangko Sentral ng Pilipinas. Two types of entities may carry out quasi-banking functions: (i) banks and (ii) non-bank financial institutions.

Section 249 of P.D. No. 69 (1972), which amended certain sections of the National Internal Revenue Code, states:

> "Quasi-banking activities shall refer to borrowing funds from twenty or more personal or corporate lenders at any one time, through the issuance, endorsement or acceptance of debt instruments of any kind other than deposits for the borrower's own accounts, or through the issuance of certificates of assignment or similar instruments, with recourse, or of repurchase agreements for purposes of relending or purchasing receivables and other similar obligations."

In short, quasi-banking is simply the issuance of deposit substitutes. The same definition was carried over in P.D. No. 71 (1972), which amended R.A. No. 337 (General Banking Act).

P.D. No. 129 (1973) allowed the Monetary Board to grant quasi-banking licenses to Investment Houses. P.D. No. 1738 (1980) defined "non-bank financial intermediary" as one authorized by the Central Bank of the Philippines to perform quasi-banking functions.

R.A. No. 7653 (The New Central Bank Act) placed non-bank financial institutions with quasi-banking functions under the regulatory oversight of the Bangko Sentral ng Pilipinas (BSP).

R.A. No. 8791 (The General Banking Law of 2000) reaffirmed the regulatory power of the BSP over quasi-banks. It also provides a stricter regime of supervision. Section 5 states, "[T]he Monetary Board may prescribe ratios, ceilings, limitations, or other forms of regulation on the different types of accounts and practices of banks and quasi-banks which shall, to the extent feasible, conform to internationally accepted standards, including of the Bank for International Settlements (BIS)."

---

[116] Cornett, M.M., and Tehranian, H., *An Examination of the Impact of the Garn-St. Germain Depository Institutions Act of 1982 on Commercial Banks and Savings and Loans*, 45 THE JOURNAL OF FINANCE 1 (1990)

The same law expressly allows banks to perform quasi-banking functions. Section 6 states, "[A]n entity authorized by the Bangko Sentral to perform universal or commercial banking functions shall likewise have the authority to engage in quasi-banking functions."

**PART VII: HOW DEPOSIT SUBSTITUTES REPLICATE THE SAFETY FEATURES OF DEPOSITS**

Unlike ordinary bank deposits, deposit substitutes are not insured by the Philippine Deposit Insurance Corporation.[117] Instead, the lender is protected through the *collateralized* nature of the transaction, or any security device that the financial institution may offer to build confidence on the product.[118] This is how deposit substitutes purportedly replicate the safety features of ordinary bank deposits.[119]

In this section, we shall discuss some of these collateral and security features in the following BSP-recognized[120] deposit substitutes: (i) repurchase agreements or "repos", (ii) promissory notes, (iii) certificates of assignment with recourse, and (iv) certificates of participation with recourse.

**A. Repos**

Through a repo, the lender and the bank enter into a sell-and-buyback agreement over a debt instrument for the purpose of replicating a collateralized loan transaction.[121] The lender acts as a buyer of the debt instrument, and gives cash funds (representing the price) to the bank.[122] The bank then delivers the debt instrument to the lender, with a promise to buy it back at maturity. Upon the arrival of the maturity date, the bank repurchases the debt instrument from the lender. The price in the original sale transaction and the buyback price are fixed in such a way that the lender obtains an amount equivalent or nearly equivalent to a debt with interest. In effect, the two parties replicate the financial incidents of a simple loan, which is structured as a sale-and-buyback agreement.[123] This is how repos replicate the features of an ordinary bank deposit, which in essence is a simple loan. This is usually worded as follows:[124]

---

[117] R.A. No. 9576
[118] *Supra* note 61.
[119] *Id.*
[120] Bangko Sentral ng Pilipinas (BSP) Manual of Regulation for Banks (MORB) dated 31 October 2015
[121] Al-Rushoud, A., and Saeed, M., *Repurchase Agreements in Financial Markets: Financial and Legal Prospective*, 22 ARAB LAW QUARTERLY 2 (2008)
[122] Money and Forex Markets: Widening Spread in Short-Term Rates, 40 ECONOMIC AND POLITICAL WEEKLY 46 (2005)
[123] *Supra* note 121.
[124] Bangko Sentral ng Pilipinas (BSP) Manual of Regulation for Banks (MORB) dated 31 October 2015

## REPURCHASE AGREEMENT

[Form template: Issue Date, Repurchase Date, consideration in PESOS, Vendor (Name of Issuer/Vendor), Vendee (Name of Vendee), sells, transfers and conveys the security(ies) described below, to be resold by Vendee and repurchased by Vendor on the repurchase date at the stated price in PESOS, subject to the terms and conditions stated on the reverse side hereof.]

**(Description of Securities)**

| Principal Debtor/s | Serial Number/s | Maturity Date/s | Face Value | Interest/Yield |
|---|---|---|---|---|
| | | | ₱ | ₱ |
| | | | | |
| TOTAL | | | | |
| | | | | |

## B. Promissory Notes

A promissory note is a "solemn acknowledgment of a debt and a formal commitment to repay it on the date and under the conditions agreed upon by the borrower and the lender."[125] It functions as a deposit substitute by creating a debt liability on the part of the bank, which undertakes to repay a principal amount with interest to the lender at maturity.[126] The bank may secure the note with collateral. This is usually worded as follows:[127]

---

[125] *Dela Rama Co. vs. Admiral United Savings Bank*, G.R. No. 154740, April 16, 2008
[126] Darmstadter, H., *Promissory Notes: Part One*, 6 BUSINESS LAW TODAY 3 (1997)
[127] Bangko Sentral ng Pilipinas (BSP) Manual of Regulation for Banks (MORB) dated 31 October 2015

```
                        PROMISSORY NOTE

                                        Issue Date    : ________, 20 ______.
                                        Maturity Date : ________, 20 ______.

FOR PESOS ________________________________ (P________), RECEIVED.
              (Present Value/Principal)

_______________________ promises to pay _______________________
  (Name of Issuer/Maker)                  (Name/Account Number of Payee)

or order, the sum of PESOS _______________________ (P________),
                            (Maturity Value/Principal & Interest)

subject to the terms and conditions on the reverse side hereof.
```

## C. Certificates of Assignment with Recourse

Through a certificate of assignment with recourse, the bank assigns, conveys, and transfers debt securities or instruments to a lender of funds, for consideration.[128] If the principal debtor under the debt security or instrument defaults, the lender may still collect from the bank. To replicate interest income, the sum of the face value and interest or yield on the debt security or instrument is higher than the consideration paid by the lender of funds to the bank. The bank foregoes some income on the debt security or instrument in exchange for liquidity, while the lender has a receivable evidenced by the certificate. This is usually worded as follows:[129]

```
              CERTIFICATE OF ASSIGNMENT WITH RECOURSE

                                        Issue Date: ___________, 20 ______.

FOR AND IN CONSIDERATION OF PESOS _______________________________

(P____________), _______________________ hereby assigns, conveys,
                      (Name of Assignor)
and transfers with recourse to _______________________ the debt of
                                   (Name of Assignee)

_______________________ to the Assignor, specifically described as follows:
  (Name of Principal Debtor)

                    (Description of Debt Securities)
```

| Principal Debtor/s | Serial Number/s | Maturity Date/s | Face Value | Interest/Yield |
|---|---|---|---|---|
|  |  |  | P | P |
|  |  |  |  |  |
|  |  | TOTAL | P | P |

and hereby undertakes that in case of default of the Principal Debtor, Assignor shall pay the face value of interest/yield on, said debt securities, subject to the terms and conditions on the reverse side hereof.

---

[128] Bangko Sentral ng Pilipinas (BSP) Manual of Regulation for Banks (MORB) dated 31 October 2015
[129] Bangko Sentral ng Pilipinas (BSP) Manual of Regulation for Banks (MORB) dated 31 October 2015

### D. Certificates of Participation with Recourse

Through a certificate of participation with recourse, the bank grants or assigns a share of its future income on debt securities or instruments to a lender, for a consideration.[130] Again, the amount of the consideration is less than the amount of the share of the lender in the future income. This is usually worded as follows:[131]

```
                    CERTIFICATE OF PARTICIPATION WITH RECOURSE

                                            Issue Date: __________, 20 ____
FOR AND IN CONSIDERATION OF PESOS _________________________________,
this certificate of participation is hereby issued to evidence the _____________
                                                                  (fraction or %)
share of _________________________________________________ in the
                        (Name of Participant)
loan/s of _____________________________ granted by/assigned to the herein issuer,

specifically described as follows:

                        (Description of Debt Securities)

| Principal Debtor/s | Serial Number/s | Maturity Date/s | Face Value | Interest/Yield |
|                    |                 |                 |     ₱      |       ₱        |
|                    |                 |                 |            |                |
|                    |                 |   TOTAL         |     ₱      |       ₱        |

The issuer shall pay, jointly and severally with the principal debtor, _____________
                                                                      (fraction or %)
share of the face value of, and the interest/yield on, said debt security(ies), subject to the terms
and conditions on the reverse side hereof.
```

## PART VIII: WHY LONG-TERM DEBT INSTRUMENTS CANNOT REPLICATE THE SAFETY FEATURES OF DEPOSITS AND DEPOSIT SUBSTITUTES

When a borrower issues a bond, note, or any debt instrument, the interest on the instrument is stipulated in advance.[132] We shall call this the "contractual" interest rate, which is usually locked over the life of the instrument as agreed upon by the parties. During the time that the lender holds the instrument—let us say for 10 years—the interest rates of other bonds, notes, and other debt instruments in the market either change or remain the same.[133] More often than not, they change due to inflation.[134] Hence, three possibilities happen while a lender holds a debt instrument with a fixed contractual interest rate:

---
[130] Bangko Sentral ng Pilipinas (BSP) Manual of Regulation for Banks (MORB) dated 31 October 2015
[131] Bangko Sentral ng Pilipinas (BSP) Manual of Regulation for Banks (MORB) dated 31 October 2015
[132] Art. 1956, New Civil Code of the Philippines
[133] Alvarez, F., Lucas, R., and Weber, W., *Interest Rates and Inflation*, 91 THE AMERICAN ECONOMIC REVIEW 2 (2001)
[134] *Id.*

1. Market interest rates are equal to the contractual interest rate;
2. Market interest rates are higher than the contractual interest rate; or
3. Market interest rates are lower than the contractual interest rate.

Let us examine the effects of these three scenarios on a bond issuance. Suppose that a borrowing entity issues a 10-year zero-coupon treasury bond with a face value of P30 billion, and an annually compounding interest rate of 10%. Since the instrument is a zero-coupon bond, the borrowing entity will pay the lender P30 billion at the time of maturity, which is the end of the 10-year period, while the lender will obtain the bond at discount (i.e. less than P30 billion), such that the return on investment will be equal to 10% interest rate, annually compounded.

The price of the bond can be obtained through the following formula:[135]

$$Bond\ Price\ = \frac{F}{(1+r)^n}$$

The face value (*F*) is P30 billion. The interest rate (*r*) is 10%, compounded annually. The time to maturity (*n*) is 10 years. Applying the formula, the price of the bond at the time of issuance is P11.57 billion. Accordingly, the lender will pay the borrower P11.57 billion in the beginning of the first year, expecting to retire the bond at maturity for P30 billion. The difference between the face value of P30 billion and bond price at time of issuance of P11.57 billion is P18.43 billion, which is called the "discount" on the bond and which represents possible investment income on the part of the lender.[136]

The lender may, of course, purchase another bond in the market from another issuer, but one crucial decision point is as follows: between two bonds having equivalent risk (called "benchmark bonds"), no rational lender will purchase that bond which yields a lower return.[137] A fair return on a bond must be equal to the market interest rate of benchmark bonds.[138] This market rate is also called the "expected return" or "required return" on the bond because it is that rate which is expected or required by an investor for a given level of risk.[139] The "fair price" of a bond is that price which is determined by the expected or required return.[140]

The fair price of the bond depends on the three scenarios on market interest rates. If the market interest rate of benchmark bonds is *equal* to the contractual interest rate of 10% at the time of issuance, then the fair price of the bond is exactly equivalent to the amount to be paid by the lender to the borrower, which is P11.57 billion.

If the market interest rate of benchmark bonds is *higher* than the contractual interest rate of 10% at the time of issuance, then the fair price of the bond is *lower* than the amount to be paid by the lender to the borrower at time of issuance. Hence, if the market interest rate is 12%, then applying the same formula as above, the fair price of the bond is P9.66 billion, which is lower than the amount paid by the lender, which is P11.57 billion. This means that the lender can obtain another bond with an equivalent risk from another borrowing entity, at a lower (i.e. more attractive) issue price.

---

[135] Damodaran, A., *Valuation: Basics*, available at: http://people.stern.nyu.edu/adamodar/pdfiles/invphiloh/valuation.pdf (last retrieved 10 March 2017)
[136] Williamson, S., *Tax-Exempt Zero-Coupon Bond Pricing*, 35 NATIONAL TAX JOURNAL 4 (1982)
[137] Duffle, D., and Stein, J., *Reforming LIBOR and Other Financial Market Benchmarks*, 29 THE JOURNAL OF ECONOMIC PERSPECTIVES 2 (2015)
[138] Ealing, C., Garcia-Feijoo, L., and Jorgensen, R.D., *Using The U.S. Treasury Market To Teach Bond Valuation And Financial Engineering*, 30 JOURNAL OF FINANCIAL EDUCATION (2004)
[139] *Id.*
[140] *Id.*

If the market interest rate of benchmark bonds is *lower* than the contractual interest rate of 10% at the time of issuance, then the fair price of the bond is *higher* than the amount to be paid by the lender to the borrower at time of issuance. If the market interest rate is 8%, then applying the same formula, the fair price of the bond is P13.9 billion, which is higher than the amount to be paid by the lender at P11.57 billion. This means that the lender can obtain a return higher than what is expected by the market from a bond of equivalent risk.

With increased transparency of data in the financial markets, investors can easily calculate if a bond is being issued at a fair price.[141] This compels the holder of a bond to sell the instrument at a price determined by market interest rates. The riskiness of a long-term debt instrument revolves around this interaction between bond prices and market rates. This is a risk that is not present in ordinary bank deposits, and by extension, in deposit substitutes.

Let us illustrate this through the same bond issuance discussed above, but this time in an environment of increasing market interest rates, which usually happens during a period of inflation. Let us suppose that the contractual interest rate of 10% is equal to the market interest rate at the time of issuance, but during the life of the bond, the market interest rate starts rising, as follows:

| Year | Market Interest Rate | Time to Maturity (beginning year) | Bond Price in PHP (beginning year) |
|---|---|---|---|
| 1 | 10% | 10 | 11,566,298,682.89 |
| 2 | 12% | 9 | 10,818,300,749.45 |
| 3 | 14% | 8 | 10,516,771,645.52 |
| 4 | 16% | 7 | 10,614,885,896.57 |
| 5 | 16% | 6 | 12,313,267,640.02 |
| 6 | 16% | 5 | 14,283,390,462.43 |
| 7 | 16% | 4 | 16,568,732,936.41 |
| 8 | 16% | 3 | 19,219,730,206.24 |
| 9 | 16% | 2 | 22,294,887,039.24 |
| 10 | 16% | 1 | 25,862,068,965.52 |
| 11 | 20% | 0 | 30,000,000,000.00 |

---

[141] Bessembinder, H., and Maxwell, W., *Markets: Transparency and the Corporate Bond Market*, 22 THE JOURNAL OF ECONOMIC PERSPECTIVES 2 (2008)

As we have stated previously, two basic safety features of deposits are capital preservation and liquidity, and these features cannot be replicated in a long-term debt instrument.

Note that there is a significant risk of capital loss, especially in the first half of the life of the bond. When market interest rate increased from 10% in year 1 to 12% in year 2, the bond price in year 2 (i.e. P10.8 billion) falls *below* the initial investment in year 1, which is P11.57 billion. This negates the idea of capital preservation or safety of principal.

Second, the risk of capital loss diminishes as the time to maturity decreases. It is only at year 5 that the investor recovers from the unrealized capital loss. At maturity, the investor is able to retire the bond at P30 billion, even with an absurdly high market interest rate of 20%. However, the fact that the investor may have to wait for a longer investment horizon before recovering his initial investment negates the idea of liquidity.

**PART IX: CRITIQUE OF THE PEACe BONDS DECISION**

In the previous section, we illustrated how a long-term debt instrument cannot have the twin features of capital preservation and liquidity. We are now ready to demonstrate this using the PEACe Bonds in *Banco De Oro, et al. vs. Republic* (2015 and 2016).

At the time of the issuance of the PEACe Bonds, they have a face value ($F$) of ₱35 billion, a contractual interest rate ($r$) of 12.75%, and a time to maturity ($n$) of 10 years.[142] Since the interest rate is compounded semi-annually, the formula for the bond price is as follows:

$$Bond\ Price = \frac{F}{(1 + \frac{r}{2})^{2n}}$$

Applying the formula, the bond has a value of ₱10.17 billion at the time of origination—i.e. RCBC paid ₱10.17 billion for a bond that the government promised to pay at ₱35 billion on the 10th year.[143] The difference between the face value of ₱35 billion and bond value of ₱10.17 billion is ₱24.83 billion. This represents the interest income to RCBC if it holds the bond until maturity.

Now suppose that RCBC holds the bond for 1 year and decides to sell it at the end of the 1st year. Suppose also that the market interest rate at the time of sale is 15%. This means that bonds having an equivalent risk profile as the PEACe Bonds pay investors a yield of 15%, which is the yield that investors would expect from similar instruments.

Applying the bond price formula and using the market interest rate of 15%, the PEACe Bonds at the end of the 1st year have a fair price of ₱9.5 billion. Since RCBC purchased the bond at ₱10.17 billion in the date of origination and sells it at ₱9.5 billion in the date of sale, RCBC suffers a loss of ₱670 million. This decrease in the price of PEACe Bonds is due to an increase in market interest rates to 15%.

---

[142] Letter of Treasurer of the Philippines to the Department of Finance dated 10 October 2011 (Re: BIR Ruling No. 370-2011 dated 07 October 2011)

[143] Letter of Department of Finance to the Bureau of Internal Revenue dated 07 October 2011 (Re: Proper Tax Treatment of the Discount or Interest Income arising from the Php 35 billion worth of 10-year zero coupon treasury bonds)

Now suppose that, in order not to realize any loss, RCBC decides not to sell the bonds, and waits until the bond price will increase. Suppose also that benchmark interest rates remain at 15% until the end of the 2nd year. Applying the formula, the PEACe Bonds sell at ₱11 billion. Since RCBC purchased the bond at ₱10.17 billion in the date of origination and sells it at ₱11 billion in the date of sale, RCBC profited by ₱830 million. However, RCBC had to wait for 2 years.

Now suppose that China Bank is the buying party which purchased the PEACe Bonds at ₱11 billion. China Bank holds the instrument for 1 year. Benchmark interest rates continue to rise, and on the 3rd year, interest rates are at 18%. China Bank sells the PEACe Bonds at the end of the 3rd year. Applying the formula, the price of the bond is ₱10.47 billion. The bank suffers a loss of ₱530 million.

If the current holder of the bond does not want to realize the loss, it will simply hold the bond for a longer period of time until the bond price will suffice to cover its acquisition cost. Notice that in both examples, investors RCBC and China Bank could not immediately withdraw the amounts of their investments within the desired time period. The fact that the holder of the instrument cannot recover the full principal amount of his investment within any time period that he chooses is incompatible with the very idea that the instrument is a deposit substitute.

## PART X: TOWARD A NEW DEFINITION OF DEPOSIT SUBSTITUTES

In light of the foregoing discussion, it is submitted that Section 22(Y) of the NIRC should be amended to exclude long-term debt instruments from the scope of deposit substitutes, with the following proposed change:

> "[A]n alternative from of obtaining funds from the public (the term 'public' means borrowing from twenty [20] or more individual or corporate lenders at any one time) other than deposits, through the issuance, endorsement, or acceptance of debt instruments ***maturing not more than 1 year from the date of origination,*** for the borrowers own account, for the purpose of relending or purchasing of receivables and other obligations, or financing their own needs or the needs of their agent or dealer."

This change rules out the possibility of classifying high-risk debt instruments, such as 10-year zero-coupon treasury notes, as deposit substitutes. The new definition would automatically subject the interest income on said instruments to the regular income tax rate, and not to 20% final tax, without prejudice to the imposition of creditable withholding tax under Revenue Regulation No. 14-2012 dated 07 November 2012.

The stability and predictability of the tax treatment of financial instruments are crucial in building investor confidence in the Philippine fiscal environment. The tax system must allow investors to plan for a long-term investment horizon and to forecast reasonable risks and returns, without being exposed to unexpected and drastic reinterpretation of tax rules. This can only be achieved if tax rules on financial instruments bear a rational connection with the nature and function of these instruments.